\documentclass[pre,aps,reprint,showkeys,floatfix]{revtex4-2}
\usepackage[utf8]{inputenc}
\usepackage{amsmath}
\usepackage{amssymb}
\usepackage{amsfonts}
\usepackage{graphicx}
\usepackage{siunitx}
\usepackage{xcolor}
\usepackage[colorlinks=true,citecolor=blue,urlcolor=red]{hyperref}
\usepackage{orcidlink}

\newcommand{\erfc}{\mathrm{erfc}}
\newcommand{\erf}{\mathrm{erf}}
\newcommand{\e}{\mathrm{e}}
\renewcommand{\d}{\mathrm{d}}

\definecolor{darkgreen}{rgb}{0,0.5,0}
\definecolor{darkred}{rgb}{0.5,0,0}
\definecolor{darkblue}{rgb}{0,0,0.5}

\begin{document}

\title{Probabilistic work extraction on a classical oscillator beyond the second law}

\author{Nicolas Barros\,\orcidlink{0009-0000-1348-7725}}\affiliation{Univ Lyon, ENS de Lyon, CNRS, Laboratoire de Physique, F-69342 Lyon, France}
\author{Sergio Ciliberto\,\orcidlink{0000-0002-4366-6094}}\affiliation{Univ Lyon, ENS de Lyon, CNRS, Laboratoire de Physique, F-69342 Lyon, France}
\author{Ludovic Bellon\,\orcidlink{0000-0002-2499-8106}}\email{ludovic.bellon@ens-lyon.fr}\affiliation{Univ Lyon, ENS de Lyon, CNRS, Laboratoire de Physique, F-69342 Lyon, France}
\date{\today}

\begin{abstract}
We demonstrate experimentally that, applying optimal protocols which drive the system between two equilibrium states characterized by a free energy difference $\Delta F$, we can maximize the probability of performing the transition between the two states with a work $W$ smaller than $\Delta F$. The second law holds only on average, resulting in the inequality $\langle W \rangle \geq \Delta F$. The experiment is performed using an underdamped oscillator evolving in a double-well potential. We show that with a suitable choice of parameters the probability of obtaining trajectories with $W \le \Delta F$ can be larger than $95\,\%$. Very fast protocols are a key feature to obtain these results, which are explained in terms of the Jarzynski equality.
\end{abstract}

\maketitle


\section{Introduction}
\label{introduction}

\begin{figure}[ht]
 \centering
 \includegraphics[width=8.5cm]{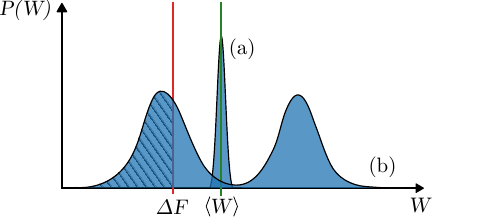}
 \caption{Probability distribution function of the work $P(W)$ during a transformation. (a) In classical macroscopic thermodynamics, fluctuations are usually Gaussian and negligible compared with the work mean value $\langle W \rangle$ (green vertical line). (b) However, in stochastic thermodynamics, such fluctuations can have a more complex distribution (such as the bimodal one sketched here), and lead to local ``violations'' of the second law: several realisations (hatched area) of a transformation can be performed with a work $W$ smaller than the free energy difference $\Delta F$ (red vertical line), even if in average $\langle W \rangle \geq \Delta F$. Note that in this sketch, the horizontal axis is rescaled between (a) and (b).\label{Fig.stoch}}
\end{figure}

Numerous experimental platforms that act on the micro and nano scales allow us to explore the laws of thermodynamics for systems with few degrees of freedom coupled with thermal baths~\cite{Ciliberto_PRX}. Typical examples are experiments in colloids~\cite{Blickle,Gavrilov_PRL_2016,Berut2012,Martinez_PRL15,Bech2014,Toyabe2010}, electric circuits~\cite{cil13a}, single electron transistors~\cite{Pekola2018,Cavina2016,Kantz_2019}, mechanical devices~\cite{Dago-2022-JStat,Joubaud2007} and single molecules~\cite{Ritort2005,Ribezzi_2019}. In such systems, fluctuations of physical quantities play a fundamental role, contrary to classical thermodynamics which mostly considers averaged quantities. Stochastic thermodynamics provides the suited framework to describe single realizations of thermodynamic transformation, and the associated work and heat distribution. In particular, the second law does not apply at the level of a single realization, and one can observe local ``violations'' due to the stochastic nature of the system, where the system can extract work, or gain free energy, at no cost to the operator and with no information feedback (illustration in Fig.~\ref{Fig.stoch})~\cite{Jarzynski_2011,Sekimoto2010,Seifert_2012,Sagawa_2014}.

In a system driven by an external control parameter $\lambda$ from an initial to a final equilibrium state resulting in a free energy difference $\Delta F$ between the two states, the probability distribution of the work performed when changing $\lambda$ is constrained by Jarzynski's equality~\cite{Jarzynski_1997} 
\begin{equation}
\langle \e^{-\beta W} \rangle = \e^{-\beta \Delta F},
\label{jar}
\end{equation}
where $\beta=1/k_BT$, with $T$ the temperature of the system and $k_B$ the Boltzmann constant. The expression of the second law can be recovered from the convexity of the exponential function: $\langle W \rangle \geq \Delta F$. We are interested here in realizations where the free energy of the system can be increased spending an amount of work $ W < \Delta F$. The free energy stored in the system can be later used, resulting in probabilistic work extraction from the device.

These realizations have been theoretically~\cite{Cavina2016} and experimentally studied~\cite{Pekola2018} with a single electron transistor with discrete energy levels, obtaining a probability of extracting work up to $\SI{65}{\%}$. Using a mechanical oscillator, we want to illustrate this behavior using a fully classical continuous system. Our goal is to experimentally obtain an arbitrary large probability of having realizations with $W<\Delta F$. To this aim, we propose a protocol that tends to the optimal work distribution predicted by Refs.~\onlinecite{Cavina2016,Kantz_2019}: it should consists in only two peaks, with the most frequent one below $\Delta F$.

\section{Experimental setup and protocol}

The experimental setup is a microcantilever, which behaves in the absence of external forces as an underdamped harmonic oscillator of stiffness $k$, resonance frequency $f_0=1200Hz$ and quality factor $Q=10$. The deflection $x$ of the cantilever is measured by interferometry~\cite{Paolino-2013}. The oscillator is in equilibrium with the surrounding air at room temperature $T$ and subject to thermal fluctuations. The variance of $x$ is $\sigma ^2 = \langle x^2 \rangle = k_B T/k \sim \SI{1}{nm^2}$. $\sigma$ is used as the unit length, and from now on all positions are expressed as dimensionless quantities $z = x / \sigma$, and energies in units of $k_B T$ (hence taking $\beta=1$ in Jarzynski's equality, Eq.~\ref{jar}). 

\begin{figure}[thb]
\centering
\includegraphics[width=8.5cm]{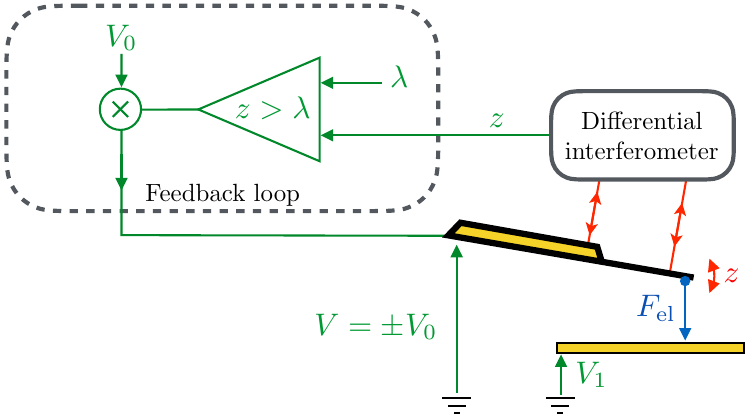}
\caption{Sketch of the experimental setup. We record the deflection $z$ of a conductive cantilever measured with high precision by a differential interferometer ~\cite{Paolino-2013}. Depending on the comparison of $z$ with a tunable threshold $\lambda$, a constant voltage $V=\pm V_0$ is applied by a fast feedback loop, creating an electrostatic force $F_\mathrm{el}$ on the cantilever towards a nearby electrode at voltage $V_1 \gg V_0$. }
\label{cantilever}
\end{figure}
\begin{figure}[thb]
\centering
\includegraphics[width=8.5cm]{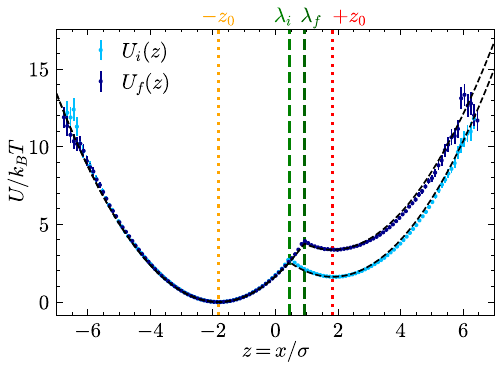}
\caption{Double well potentials. The potential energy of the cantilever is the juxtaposition of two harmonic wells centered in $\pm z_0=\pm 1.8$, the switch between the two occurring when $z=\lambda$. The two examples plotted here correspond to $\lambda_i = 0.45$ and $\lambda_f = 0.92$. The measured potential $U(z)$ data is inferred from Eq.~\ref{eq:Boltzmann} and the pdf of the experimental positions $z$ during a long acquisition. The black dashed lines are best fit to Eq.~\ref{eqpotentiel}, leading to the aforementioned values of $z_0$ and $\lambda$. The transformation we apply corresponds to a step of $\lambda$ from $\lambda_i$ to $\lambda_f$, leaving the lower well untouched while raising the upper one by $\Delta U=2 (\lambda_f-\lambda_i) x_0$.}
\label{Fig.potential}
\end{figure}

To tune the potential experienced by the resonator, we use an electrostatic force acting on the cantilever. A time-dependent virtual potential can be created by a fast feedback loop~\cite{Dago-2022-JStat,Dago2023} that adjusts the voltage $V$ applied to the cantilever (thus the electrostatic force) depending on the measured position $z$ (see Fig.~\ref{cantilever}). We implement the following simple rule: $V=-V_0$ if the position is below a threshold $\lambda$, and $V=+V_0$ otherwise. This creates an asymmetric double well, illustrated in Fig.~\ref{Fig.potential}. Two parameters are available to tune the potential: $\lambda$ sets the barrier position and $V_0$ sets the centers of the two wells in $\pm z_0 \propto \pm V_0$. 
Theoretically, the potential energy constructed by this feedback is:
\begin{equation}
\begin{split}
U(z,\lambda,z_0) = & \frac{1}{2} \big(z-S(z-\lambda)z_0\big)^2 \\
 & + \lambda z_0\big(S(z-\lambda) +S(\lambda)\big),
\end{split}
\label{eqpotentiel}
\end{equation}
where $S$ is the sign function: $S(z-\lambda) = - 1$ if $z < \lambda$ and $S(z-\lambda) = 1$ if $ z > \lambda$. To illustrate the validity of this model for $U$, we record a long time trace of the position in a static potential to evaluate the probability distribution function (pdf) of the position $P(z)$. We then reconstruct the double well using Boltzmann's prescription at equilibrium:
\begin{subequations} \label{eq:Boltzmann}
\begin{align}
P(z) & = \frac{1}{Z(\lambda,z_0)} \e^{-U(z,\lambda,z_0)},\label{eqP(z)}\\
Z(\lambda,z_0) & = \int_{-\infty}^{+\infty} \e^{-U(z,\lambda,z_0)} \d z,\label{eqZ}
\end{align}
\end{subequations}
with $Z$ the partition function. As plotted in Fig.~\ref{Fig.potential}, the model is an excellent description of the effective potential evaluated through $U(z)=-\ln [P(z)/P(-S(\lambda)z_0)]$.

\begin{figure}[!b]
\centering
\includegraphics[width=8.5cm]{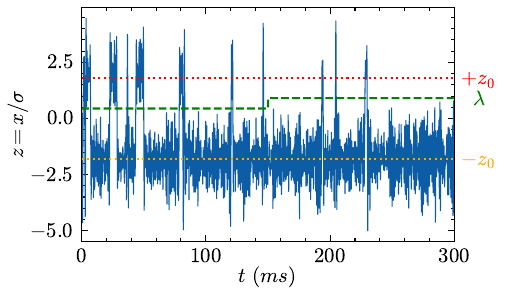}
\caption{Trajectory of an underdamped oscillator evolving in a time-dependent asymmetric double-well potential. The position $z$ is expressed in units of standard deviation $\sigma$ in a single well. The center of the wells is kept constant at $\pm z_0 = \pm 1.8$, but the commutation threshold $\lambda$ is changed in less than $\SI{5}{\mu s}$ from $\lambda_i = 0.45$ to $\lambda_f = 0.92$ at $t=\SI{150}{ms}$. }
\label{Fig.trace} 
\end{figure}

To observe local ``violations'' of the second law, we design the following protocol between an initial state (all quantities labeled by the subscript $_i$) and a final state (subscript $_f$). First, the cantilever evolves at equilibrium in an initial double-well potential $U_i(z)=U(z,\lambda_i,z_0)$. Then, we instantaneously increase the threshold $\lambda$ between the two wells, from $\lambda_i$ to $\lambda_f$. Finally, the cantilever is left in the final potential $U_f(z)=U(z,\lambda_f,z_0)$. The well centers are left unchanged during the protocol, $z_0$ keeping the same value. An example of the time trace of one realisation of such protocol is plotted in Fig.~\ref{Fig.trace}.

Initial and final potentials are represented in Fig.~\ref{Fig.potential}. Each potential is measured from the equilibrium pdf $P(z)$ in the initial and final state, then fitted with very good accuracy using Eq.~\ref{eqpotentiel}. The experimental values of the parameters $\lambda_i$, $\lambda_f$ and $z_0$ are deduced from the fit. Potential wise, the protocol amounts to leaving the lower well unaffected while raising the upper one. The work $W$ performed will be $0$ most of the time (anytime the system is in the lower well, which is likely), while the free energy difference is positive (since $U(x)$ is globally increasing): the probability of observing $W<\Delta F$ should thus be high. 

The values of $z_0 = 1.8$ and $\lambda_f = 0.92$ are the same for all experiments. We tune the initial threshold $\lambda_i$ with values going from 0 to 0.83. For each one of them, the protocol detailed in Fig.~\ref{Fig.trace} is repeated around $N \sim 2500$ times. This allows us to obtain a good enough statistics to estimate the pdf $P(z)$, and further on the work distribution.

\section{Analysis}

Following the classical convention of stochastic thermodynamics~\cite{Sekimoto2010}, we define the work $W$ done by the operator through the variations of the external parameters tuning the potential $U(z,\lambda,z_0)$: 
\begin{equation}
W = \int \frac{\partial U}{\partial \lambda }\dot \lambda \d t.
\label{eqworksimp}
\end{equation}
Since $z_0$ is kept constant in all protocols, we do not consider the term in $\dot z_0$ which is always zero here. $W$ is computed using the recorded trajectories $z(t)$ and Eq.~\ref{eqpotentiel}. Since the variation of $\lambda(t)$ is a step function, the work can be equivalently computed as $W=U_f(z)-U_i(z)$ using the value of $z$ at the moment of the switch. From this expression we can also infer the theoretical expectation for $\langle W \rangle$:
\begin{subequations}
\begin{align}
\langle W \rangle & = \int_{-\infty}^{+\infty} \big[U_f(z)-U_i(z)\big] \frac{1}{Z_i}\e^{-U_i(z)} \d z, \\
& = \frac{2 z_0 \e^{-2 \lambda_i z_0}}{Z_i}\big[f(\lambda_f-z_0)-f(\lambda_i-z_0)\big],
\end{align}
\end{subequations}
where $f$ is defined by
\begin{equation}
 f(u)=\sqrt{\pi/2} \, u\, \erfc\left(-u/\sqrt{2}\right)-\e^{-u^2/2}.
\end{equation}
$Z_i=Z(\lambda_i,z_0)$ is the partition function in the initial state, which can be computed from Eq.~\ref{eqZ} as:
\begin{equation}
\begin{split}
Z(\lambda,z_0) = \sqrt{\frac{\pi}{2}}\bigg[2 & -\erfc\bigg(\frac{\lambda+z_0}{\sqrt{2}}\bigg)\\
& +\e^{-2\lambda z_0}\erfc\bigg(\frac{\lambda-z_0}{\sqrt{2}}\bigg)\bigg].
\end{split}
\end{equation}
As shown in Appendix \ref{AppendixPW}, the full distribution of the work $P(W)$ can actually be computed in this playground.

The free energy difference in the system $\Delta F = F_f - F_i$ can be computed in two different ways. For a given protocol, the work distribution obeys Jarzynski's equality (Eq.~\ref{jar}), thus $\Delta F = - \ln \langle e^{-W} \rangle$ can be deduced from the experimental work distribution with good enough statistics. The second approach is to use the partition function $Z$ at equilibrium: since the free energy of the system is $F=-\ln (Z)$, the free energy difference $\Delta F = \ln(Z_i/Z_f)$ can be theoretically directly calculated from the parameters of the initial and final potentials.

\begin{figure}[b]
\centering
\includegraphics[width=8.5cm]{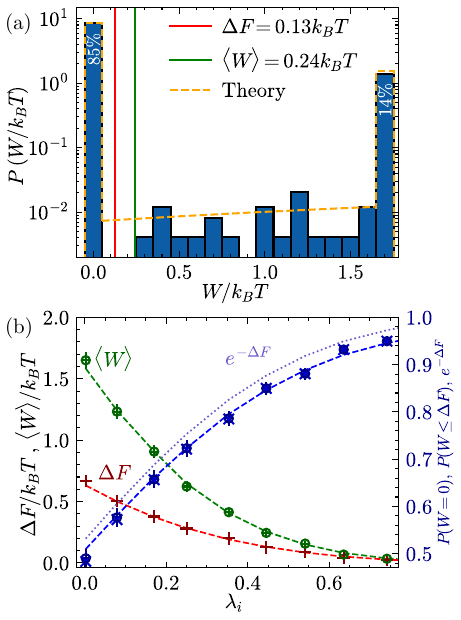}
\caption{ (a) Work distribution for $N=2450$ protocols, corresponding to the protocol in Fig.~\ref{Fig.potential} ($\lambda_i = 0.45$, $\lambda_f = 0.92$). We also report the mean work $\langle W \rangle = 0.24 \pm 0.02$ and free energy difference $\Delta F = 0.13 \pm 0.01$ computed using Jarzynski's equality. We observe a transient violation of the second law in 85\% of the protocols (note the vertical log scale), when $W=0$. The other values of the work are mostly $W=W_{\max}$ (right peak with 14\% of the occurrences). The last 1\% of the trajectories is distributed between the two extreme values. Although poorly sampled, it is compatible with the theoretical expectation (dashed orange curve, including the two delta peaks plotted here using a bin width of $0.1$, from Eqs.~\ref{EqPW0}, \ref{EqPWmax} and \ref{EqPWint} in Appendix \ref{AppendixPW}).
(b) Mean work $\langle W \rangle$ ({\large\color{darkgreen}$\circ$}, top curve) and free energy difference $\Delta F$ ({\color{darkred}$+$}, bottom curve) and for all initial conditions $\lambda_i$. Experiment (markers with error bars corresponding to one standard deviation of the statistical uncertainties) and theory (dashed lines) are in excellent agreement. The second law of thermodynamics $\langle W \rangle \geq \Delta F$ always holds, though the probability of observing the contrary on the single trajectory reaches $\SI{95}{\%}$ for the highest values of $\lambda_i$ ({\large\color{darkblue}$\circ$} marker, experiment, and dashed blue line, theory, right vertical scale). The probability of observing a zero work $P(W=0)$ ({\color{darkblue}$\times$}) coincide with the one of transiently violating the second law, and is upper bounded by $\e^{-\Delta F}$ (dotted blue line, right vertical scale), as predicted by Eq.~\ref{eq:Ple0}~\cite{Jarzynski_2008}.}
\label{FigPWDF}
\end{figure}

\section{Results}

The protocol is repeated $N \sim 2500$ times for 10 values of $\lambda_i$ ranging from 0 to 0.83. For each $\lambda_i$, we compute the work distribution. An example for $\lambda_i = 0.45$ is plotted in Fig.~\ref{FigPWDF}(a), where we also report the mean work $\langle W \rangle$ performed by the driving, and the difference in free energy $\Delta F$. The work distribution consists mostly of two narrow peaks. The first one, for $W=0$, corresponds to the cases where the cantilever is in the lower well ($z<\lambda_i$) of the potential during the step of $\lambda$. Since there is no change in the lower part of the potential, there is no energy cost. The second peak comes from the cases where $z>\lambda_f$: in this area, the potential is shifted by $\Delta U = W_{\max} = 2z_0(\lambda_f-\lambda_i)$, and driving the parameter $\lambda$ implies this amount of work. We also observe some intermediate values, corresponding to $\lambda_i<z<\lambda_f$: the cantilever is initially in the upper well but ends in the lower one due to the step in $\lambda$, resulting in an intermediate energy change $\Delta U = 2z_0(z-\lambda_i)$, where $z$ is the value of the deflection at the time of the switch.

As illustrated in Fig.~\ref{FigPWDF}(b), the second law is always satisfied ($ \langle W \rangle > \Delta F$), but we manage to observe an almost arbitrary large proportion of transient violations ($W < \Delta F$). Indeed, by tuning the initial asymmetry of the potential, we can increase the probability of being in the lower well ($z<\lambda_i$) during the switch. The probability of observing a transient violation of the second law $P( W < \Delta F )$ can thus be arbitrary large. With our choice of parameters, for values of $\lambda_i$ very close to the final value of $\lambda_f$, we manage to measure values of $P( W < \Delta F )$ of 95\%. Our experimental results are in excellent agreement with the theoretical expectation, whose analytical expression is given by Eq.~\ref{Eq.PWlessthanDF} in Appendix \ref{AppendixPW}
However, the free energy $\Delta F$ decreases when increasing $\lambda_i$: work extraction is more and more likely, but the gain with respect to $\Delta F$ decreases.

Another theoretical result that we can probe with our experiment is the inequality~\cite{Jarzynski_2008}:
\begin{equation} \label{eq:Ple0}
P (W \le 0) \le \e^{-\Delta F}.
\end{equation}
In our case, $P (W \le 0)=P (W = 0)$ since $W$ cannot be negative. Moreover, this probability is very close to $P (W \le \Delta F)$ since the work distribution consists mainly of two peaks: the one in $0$ and the other in $W_{\max}$, above $\Delta F$. Some trajectories present a work in between the two peaks, but they are infrequent and poorly sampled in our experiment. The two probabilities $P (W \le \Delta F)$ and $P (W = 0)$ therefore coincide within statistical uncertainties, as shown in Fig.~\ref{FigPWDF}(b). In this same figure, we plot the upper bound given by Eq.~\ref{eq:Ple0}, which is indeed confirmed in our experiment, and close to be saturated.

It is interesting to notice that in our system the dissipation (the quality factor $Q$) can be changed by controlling the pressure of the air surrounding the cantilever. However as we start from equilibrium and the work corresponds to an instantaneous change of the potentials, the dynamics of the system and the quality factor have no influence on the results plotted in Fig.~\ref{FigPWDF}: the work distribution and $\Delta F$ would be the same in an overdamped system.

\section{Conclusion}

We have shown, using a fully classical continuous mechanical system, how we can observe an arbitrary large number of apparent ``violations'' of the second law of thermodynamics, while being consistent with the rules of stochastic thermodynamics. We show a clear trade-off between this probability and the free energy gained during those events. We reach the theoretical limit of $95\%$ probability (for our set of parameters) of having anomalous trajectories with $W<\Delta F$. This result is made possible by the specific way in which the potential is driven during the protocol: the center of the two wells does not change and only the minimum of the upper well is raised. We show in Appendix \ref{AppendixB} that for constant stiffness wells, this protocol is the most efficient. Indeed it produces a work probability distribution with mainly two Dirac functions that matches the optimal distribution described in Ref.~\cite{Cavina2016,Kantz_2019} to maximize the work extraction probability. One Dirac peak is centered in zero and corresponds to the trajectories that start in the lower well. The other is centered to a positive value of the work and corresponds to the trajectories starting on the upper well. In our experiment the relative amplitude of the two peaks, which determines the amount of ``anomalous'' trajectories, can be tuned by changing the minimum positions of the wells through the value of $\lambda$. In this way we have transformed for the transition properties a continuous classical system to a two levels system using a protocol similar to the one of Ref.~\onlinecite{Kantz_2019} which requires only a quench. In this context, the fast switch between the initial and final state is a key ingredient of the protocol, again as proposed in Refs.~\onlinecite{Cavina2016,Kantz_2019}. However, we do not use slow ramps as proposed in Ref.~\onlinecite{Cavina2016} and experimentally applied in Ref.~\onlinecite{Pekola2018}. Indeed a slow ramp would broaden the peaks of the work distribution, resulting in a situation similar to Fig.~\ref{Fig.stoch} and those described by Fig.~\ref{Fig.P(W)} in Appendix \ref{AppendixB}. For symmetric distributions as well, where the mean and the median are equal, observing $W<\Delta F$ is unlikely: the probability is smaller than $\SI{50}{\%}$, this limit being reached in the reversible limit.

Let us conclude by saying that in spite of the fact there is an energy gain for $95\%$ of the trajectories, the total mean remains positive. In order to use this energy surplus one should introduce a demon which selects the good trajectories, i.e. those starting on the lower well. Of course such a demon would need some energy to elaborate the information gathered from the dynamics, and the second law of thermodynamics will still hold overall. However the advantage of this demon is that this energy loss could be spent remotely (in space or time) with respect to that of the standard operation of the system. It therefore decouples the transformation of the system from the necessary energy consumption that can be spent elsewhere or at some other time. This could, for instance, allow for a chemical reaction to stay cool during an exothermic transformation, or model some enzyme behavior in biological processes.

\null

The data supporting this study are openly available in Ref.~\onlinecite{Barros-Zenodo-2024}.

\acknowledgments
This work has been supported by project ANR-22-CE42-0022. We thank C. Jarzynski for enlightening physical discussions.

\appendix
\section*{Appendix}

\section{$P(W)$ when changing only $\lambda$} \label{AppendixPW}

\begin{figure}[t]
\centering
\includegraphics{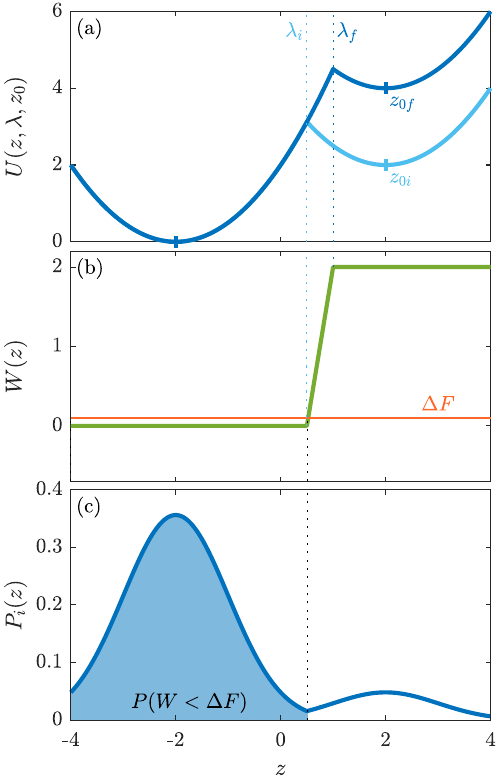}
\caption{(a) Examples of the double well initial ($z_{0}=2$, $\lambda_i=0.5$, light blue) and final ($z_{0}=2$, $\lambda_f=1$, dark blue) potentials. (b) The work $W$ performed on the system during an instantaneous jump between the potentials depends is linear in $z$ by piece. (c) The probability $P(W<\Delta F)$ is simply the area under the pdf $P_i(z)$ in the corresponding $z$ range.}
\label{Fig.potential0}
\end{figure}

The work performed during the instantaneous switch between the two bi-quadratic wells depends on the value of $z$ during the switch (see Fig.~\ref{Fig.potential0}):
\begin{subequations}
\begin{align}
    W(z) &= 0 &&\text{if } z\le \lambda_i\\
    W(z) &= 2 z_0 (z-\lambda_i) &&\text{if } \lambda_i<z<\lambda_f,\\
    W(z) &= W_{\max} = 2 z_0 (\lambda_f-\lambda_i) &&\text{if } z \ge \lambda_f.
\end{align}
\end{subequations}
Knowing the pdf of the initial position $P_i(z)=\exp[-U_i(z))]/Z_i$ from the equilibrium in the initial potential $U_i(z)$, it is straightforward to compute 
\begin{align}
 P(W=0) &= \int_{-\infty}^{\lambda_i} \frac{1}{Z_i} \e^{-U_i(z)} \d z \\
 &= \frac{1}{Z_i}\sqrt{\frac{\pi}{2}}\left[2-\erfc\left(\frac{\lambda_i+z_0}{\sqrt{2}}\right)\right],\label{EqPW0}\\
 P(W=W_{\max}) &= \int_{\lambda_f}^{+\infty} \frac{1}{Z_i} \e^{-U_i(z)} \d z \\
 &= \frac{1}{Z_i}\e^{-2\lambda_i z_0}\sqrt{\frac{\pi}{2}}\erfc\left(\frac{\lambda_f-z_0}{\sqrt{2}}\right).\label{EqPWmax}
\end{align}
For the intermediate values of $W$, the pdf writes:
\begin{align}
 P(W) &= P_i(z) \left|\frac{\d z}{\d W}\right| \\
 &=\frac{1}{Z_i}\e^{-2\lambda_i z_0}\e^{-\frac{1}{2}\left(W/2z_0+\lambda_i-z_0\right)^2}\frac{1}{2z_0}. \label{EqPWint}
\end{align}
An example of this pdf is plotted in Fig.~\ref{FigPWDF}(a). The two peaks match the prediction of Eqs.~\ref{EqPW0} and \ref{EqPWmax}. The intermediate values of the work are infrequent and poorly sampled, but compatible with Eq.~\ref{EqPWint}.

From the knowledge of the pdf $P(W)$, it is easy to estimate the probability of observing a local violation of the second principle:
\begin{align}
    P( W < \Delta F ) &= \int_{0}^{\Delta F} P(W) \d W\\
    &= P( W = 0 ) + \frac{1}{Z_i}\sqrt{\frac{\pi}{2}} \e^{-2\lambda_i z_0}\times \label{Eq.PWlessthanDF} \\
\begin{split}
 \bigg[\erf\bigg(\frac{\Delta F}{2\sqrt{2}z_0}+\frac{\lambda_i-z_0}{\sqrt{2}}\bigg)-\erf\bigg(\frac{\lambda_i-z_0}{\sqrt{2}}\bigg)\bigg]. \nonumber
    \end{split}
\end{align}
This prediction, plotted in Fig.~\ref{FigPWDF}(b), presents an excellent agreement with the experiment.

\section{$P(W)$ when changing both $\lambda$ and $z_0$} \label{AppendixB}

\begin{figure}[b]
\centering
\includegraphics{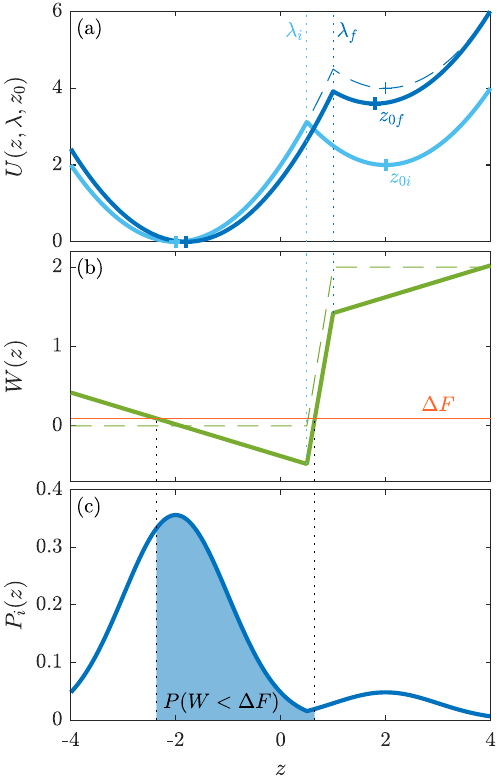}
\caption{(a) Double well initial ($z_{0i}=2$, $\lambda_i=0.5$, light blue) and final ($z_{0f}=1.8$, $\lambda_f=1$, dark blue) potentials. The case of constant $z_0$ (same as Fig.~\ref{Fig.potential0}) is plotted with a dashed line. (b) The work $W$ performed on the system during an instantaneous jump between the potentials depends on $z$ and is linear by piece. The range of $z$ for which $W<\Delta F$ is smaller for $z_{0f}=1.8$ (plain line) than for $z_{0f}=2$ (dashed line). (c) The probability $P(W<\Delta F)$ is directly read from the area under the pdf $P(z)$ in the corresponding $z$ range. The change between $z_{0i}$ and $z_{0f}$ excludes the left part of the lower well from this probability, lowering substantially the work extraction occurrences.}
\label{Fig.potential2}
\end{figure}

We use the same bi-quadratic potential energy as previously, but we now allow a transformation where both $\lambda$ and $z_0$ are suddenly changed from ($\lambda_i$, $z_{0i}$) to ($\lambda_f$, $z_{0f}$). An example of such initial and final potentials is plotted in Fig.~\ref{Fig.potential2}. The work $W$ performed during a single trajectory depends on the value of $z$ at the instant of the switch and writes:
\begin{equation}
    W(z)=a(z) z + 2b(z) + \frac{z_{0f}^2-z_{0i}^2}{2},
\end{equation}
where
\begin{align}
    a &= z_{0f}-z_{0i}, &&b = 0 &&\text{if } z\le\lambda_i,\lambda_f,\nonumber\\
    a &= z_{0f}+z_{0i}, &&b = -\lambda_i z_{0i} &&\text{if } \lambda_i<z<\lambda_f,\nonumber\\
    a &= -z_{0f}-z_{0i}, &&b = \lambda_f z_{0f} &&\text{if } \lambda_f<z<\lambda_i,\nonumber\\
    a &= -z_{0f}+z_{0i}, &&b = \lambda_f z_{0f}-\lambda_i z_{0i} &&\text{if } z\ge\lambda_i,\lambda_f.\nonumber
\end{align}
It is thus linear by parts, as the example plotted in Fig.~\ref{Fig.potential2}(b). It is easy to see in this figure that the probability of getting $W<\Delta F$ is smaller when $z_{0f}\ne z_{0i}$, since the range of $z$ for which this inequality holds is reduced with respect to the fixed $z_0$ case. The pdf of $W$ can be computed from the one in $z$, making sure to sum other all values of $z$ that give the save value of the work $W$ (since $W(z)$ is bijective only by part):
\begin{align}
    P(W)&=\sum_{z|W(z)=W} P(z)\left|\frac{\d z}{\d W}\right|\label{eqPW}\\
    &= \sum_{z|W(z)=W} \frac{1}{|a(z)|Z_i}\e^{-U_i(z)}.
\end{align}
In the example of Fig.~\ref{Fig.potential2}, each value of $W$ corresponds to two values of $z$, one above and one below $\lambda_i$, as long as $W>\min(W)=W(\lambda_i)=(z_{0f}-z_{0i})\lambda_i+\frac{1}{2}(z_{0f}^2-z_{0i}^2)$.

\begin{figure}[b]
\centering
\includegraphics{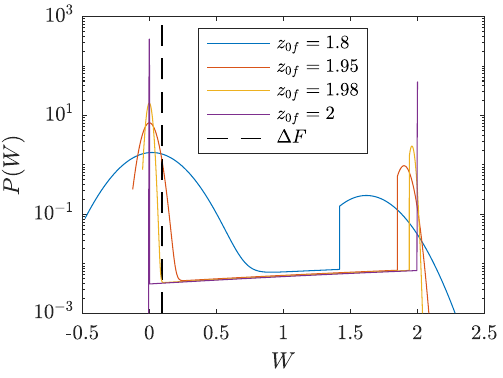}
\caption{Probability distribution function $P(W)$ of the work extracted during a sudden jump from  $z_{0i}=2$, $\lambda_i=0.5$ to $z_{0f}=1.8$ to $2$, $\lambda_f=1$. The corresponding $\Delta F$ is plotted with vertical black dashed lines, which are very close and all superpose in this graph. When we depart from $z_{0f}=z_{0i}$, the delta peaks broaden and the lower one crosses the $\Delta F$ threshold: work extraction becomes less likely in this case.}
\label{Fig.P(W)} 
\end{figure}

A few examples of $P(W)$ are plotted in Fig.~\ref{Fig.P(W)}. We see that as soon as $z_{0f}\ne z_{0i}$, the delta function parts of the pdf widen, as it becomes a truncated Gaussian of variance $(z_{0f}-z_{0i})^2$. The low work peak therefore spills into the area $W>\Delta F$. All those opportunities to gain some work with respect to $\Delta F$ are lost, as we depart from the optimal two Dirac distribution.

\begin{figure}[t]
\centering
\includegraphics{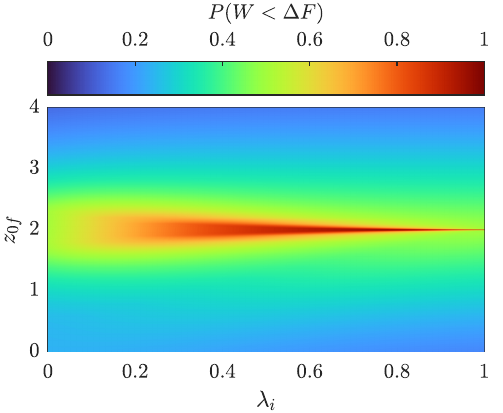}
\caption{Probability of extracting a work $W<\Delta F$ during a sudden jump between ($z_{0i}=2$, $\lambda_i=0$ to $1$) and ($z_{0f}=0$ to $4$, $\lambda_f=1$). The most frequent local violation of the second law are obtained for $z_{0f}=z_{0i}$.}
\label{Figmap} 
\end{figure}

A larger exploration of the probability of local violation of the second principle while scanning the values of $\lambda_i$ and $x_{0f}$ is plotted in Fig.~\ref{Figmap} in the form of a 2D heat map. It is clear from this figure that the highest probabilities correspond to $z_{0f} = z_{0i}$.

In our current implementation of the protocol, we use only bi-quadratic potentials of constant stiffness controlled by parameters $z_0$ and $\lambda$. Keeping $z_0$ constant during the transformation is in this case the right choice to maximize probabilistic work extraction.

\section{$P(W)$ for fast arbitrary protocols}

The approach to compute the pdf of the work from the initial potential $U_i(z)$ to the final one $U_f(z)$ in an instantaneous step is not limited to the bi-quadratic potential and can be extended to any shape. The same recipe should be applied: compute the work $W(z)=U_f(z)-U_i(z)$, split the curve into monotonous (thus bijective) parts, and apply Eq.~\ref{eqPW} to infer $P(W)$. The threshold for work extraction beyond the second principle is computed likewise with $\Delta F = \ln (Z_i/Z_f)$, with $Z=\int \exp[-U(z)]\d z$ being the partition function.

Broadening the shapes available to design the potential energy landscape could then open possibilities to optimize work extraction in more general cases. An interesting model case would be decreasing the stiffness of the lower well while raising the upper one: an adequate tuning of the parameters would yield $\Delta F = 0$, and all trajectories corresponding to $W<\Delta F=0$ to actual work extraction thanks to the expansion of the lower well. Other more complex energy landscapes could also correspond to practical problems in biophysical transformations or chemical reactions.

\bibliography{HighProba}

\end{document}